\documentclass[aps,prb,twocolumn,superscriptaddress,longbibliography]{revtex4-2}

\usepackage{graphicx}
\usepackage{amsmath,amssymb}
\usepackage{bm}
\usepackage{siunitx}
\usepackage{float}
\usepackage{subfig}
\usepackage{ragged2e}

\makeatletter
\long\def\@makecaption#1#2{%
	\par
	\begingroup
	\small
	\justifying
	#1. #2\par
	\endgroup
}
\makeatother

\begin{document}
	
	\title{Low-frequency noise as a probe of microscopic disorder in CVD-grown graphene}
	
	\author{Jagadis Prasad Nayak}
	\email{jagadisnayak07@gmail.com}
	
	\author{Smrutirekha Sahoo}
	
	\author{Shreya Barman}
	
	\author{Gopi Nath Daptary}
	\email{daptarygn@nitrkl.ac.in}
	
	\affiliation{Department of Physics and Astronomy, National Institute of Technology Rourkela, Rourkela - 769008, India}
	
	\begin{abstract}
		
		We report a detailed investigation of low-frequency resistance fluctuations (1/f noise) in chemical vapor deposition (CVD) grown graphene. Systematic measurements reveal that the magnitude of 1/f noise in CVD-grown graphene is significantly higher by several orders of magnitude than that typically observed in exfoliated single-crystal graphene. This enhancement is attributed to structural imperfections such as grain boundaries and defect states within the polycrystalline film. Detailed analysis of the temperature dependence of the noise demonstrates that the resistance fluctuations arise from thermally activated dynamics of localized defects. These results provide key insights into the microscopic mechanism of noise in scalable graphene films and highlight the role of defect engineering in optimizing graphene for large-scale electronic applications. Our findings establish low-frequency noise as a sensitive probe of microscopic disorder in CVD graphene, providing a practical pathway for assessing material quality in scalable electronic technologies.
		
	\end{abstract}
	
	\keywords{CVD-grown Graphene, low-frequency electrical noise, defect-mediated transport, grain boundary effects, thermally activated fluctuation dynamics}
	
	\maketitle
	
\section{INTRODUCTION}
Graphene is a single layer of carbon atoms arranged in a two-dimensional honeycomb lattice \cite{ref1}. Because of this unique atomic arrangement, the electrons in graphene behave very differently from those in conventional materials. Instead of exhibiting the usual parabolic energy–momentum relationship, the low-energy electronic dispersion in graphene is linear, closely resembling the physics of relativistic particles described by the Dirac equation. As a result, charge carriers in graphene behave as massless Dirac fermions \cite{ref2,ref3}, and can be either electrons or holes depending on the applied gate voltage or doping. A key feature of graphene’s band structure is the Dirac point, the energy level at which the conduction and valence bands meet. In ideal graphene, this point lies exactly at the Fermi level and corresponds to zero carrier density. In momentum space, these Dirac points occur at the high-symmetry $K$ and $K'$ corners of the Brillouin zone \cite{ref4}.

In high-quality graphene devices such as those fabricated by mechanical exfoliation or suspended over a substrate the Dirac point appears extremely sharp \cite{ref5,ref6}. This sharpness reflects the minimal presence of disorder and charged impurities, allowing researchers to probe the intrinsic properties of graphene. Such exfoliated samples have played a major role in fundamental research and in revealing many remarkable quantum transport phenomena that make graphene exceptional \cite{ref5,ref7,ref8}. However, despite their superior quality, exfoliated samples are typically small only a few micrometres in lateral size which limits their suitability for real-world technologies. Applications such as flexible electronics, sensors, large-area interconnects, and transparent electrodes require scalable and industrially compatible graphene production. For this reason, large-scale fabrication techniques such as chemical vapor deposition (CVD) have become the focus of technological development \cite{ref9,ref10}. However, large-area graphene films often contain greater structural and chemical disorder compared with exfoliated samples due to grain boundaries, wrinkles, transfer residues, vacancies, and substrate interactions. These imperfections can strongly influence the electronic transport properties of graphene \cite{ref11,ref12,ref13,ref14,ref15,ref16}. Recent studies have also shown that such disorder affects the behavior of graphene when integrated with superconducting materials, leading to unusual electronic properties in graphene-based hybrid systems \cite{ref17,ref18}. Understanding the microscopic mechanisms governing charge carrier dynamics in large-area graphene is therefore essential, both for improving material quality and for enabling reliable device performance.

Low-frequency electrical noise (commonly known as $1/f$ noise) has emerged as a powerful experimental probe for studying microscopic disorder and carrier dynamics in two-dimensional materials \cite{ref19,ref20,ref21,ref22,ref23,ref24}. In graphene and few-layer graphene devices, measurements of $1/f$ noise as a function of gate voltage reveal contributions from defect- and trap-mediated fluctuations, carrier number and mobility fluctuations, and screening effects that depend on the thickness of the material. These studies demonstrate that noise spectroscopy is a valuable tool for probing disorder and transport mechanisms in atomically thin conductors \cite{ref20,ref21,ref25,ref26,ref27,ref28,ref29}. Similar investigations in transition-metal dichalcogenides (TMDs), such as single-layer and bilayer MoS$_2$, reveal $1/f$ noise governed by mobility or carrier-number fluctuations influenced by adsorbates, trap states, temperature, and device environment \cite{ref19,ref30,ref31,ref32}. Studies in other emerging TMD materials such as MoTe$_2$ further indicate that variable-range hopping and trap-related mechanisms contribute significantly to the observed noise characteristics, demonstrating that noise measurements provide insights into transport processes beyond conventional DC characterization \cite{ref33}.Noise measurements in quasi-one-dimensional Weyl semimetal $(\mathrm{TaSe}_4)_2\mathrm{I}$ nanoribbons exhibit clear $1/f$ behavior with strong enhancement and Lorentzian features near the charge-density-wave (Peierls) transition, illustrating that noise spectroscopy can sensitively probe collective electronic phases in reduced-dimensional topological systems \cite{ref34}. Extensions of $1/f$ noise analysis to oxide-based quasi-two-dimensional electron systems, including LaAlO$_3$/SrTiO$_3$ and related complex oxide heterostructures, show that charge trapping, interfacial disorder, multiband transport, and phase fluctuations dominate the low-frequency noise response, providing insight into defect distributions and electron dynamics in these interfaces \cite{ref35,ref36,ref37,ref38}. Collectively, these studies demonstrate the universality of $1/f$ noise as a powerful diagnostic tool for probing microscopic disorder, carrier dynamics, and phase fluctuations in low-dimensional systems.

In this work, we present a detailed investigation of the temperature-dependent resistance and low-frequency resistance fluctuations in large-area graphene synthesized via chemical vapor deposition. We find that the noise magnitude in polycrystalline CVD-grown graphene is significantly higher compared to exfoliated single-crystal graphene devices. Our analysis indicates that the observed $1/f$ resistance noise originates from thermally activated fluctuations associated with local defect states within the graphene film.

\begin{figure}[H]
	\includegraphics[width=0.35\textwidth]{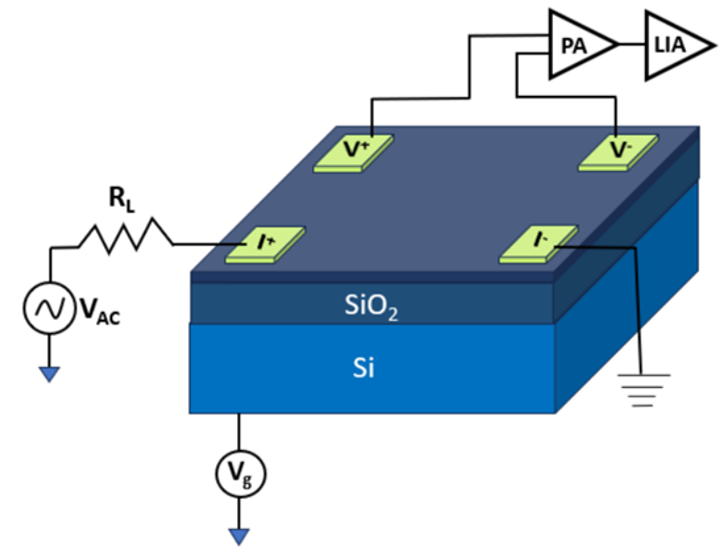}
	\caption{Schematic of the CVD-grown graphene device.}
	\label{fig:fig1}
\end{figure}

	\section{Results and Discussion}
\begin{figure}[H]
	\includegraphics[width=0.48\textwidth]{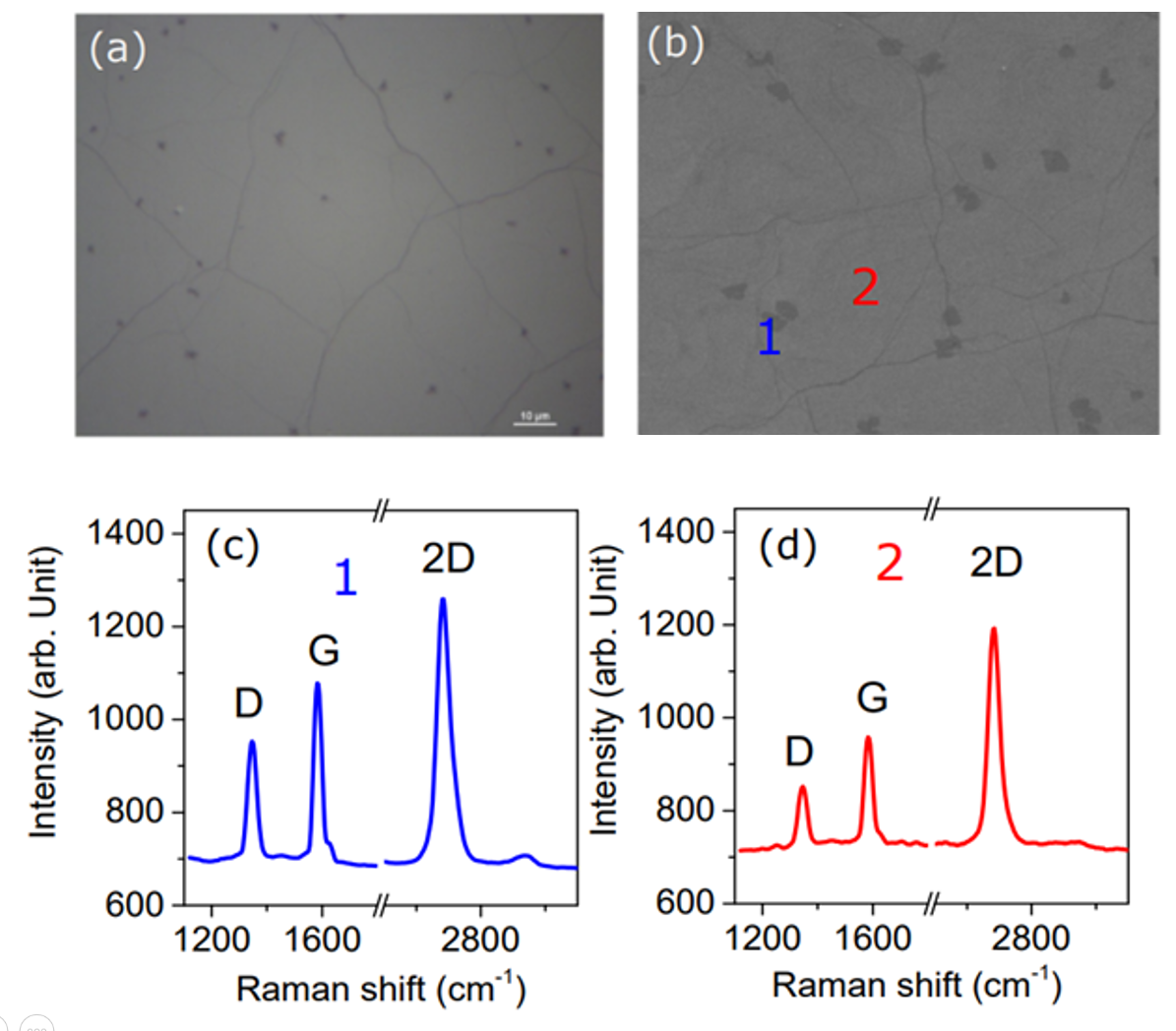}
	\caption{Characterization of CVD-grown graphene. (a) Optical microscope image and (b) SEM image of the graphene film. (c) and (d) Raman spectra measured at the grain boundary (location 1 in (b)) and within a single grain (location 2 in (b)), respectively, highlighting the variation in spectral features between the two regions}
	\label{fig:fig2}
\end{figure}

	The structural properties of the CVD-grown graphene were characterized using optical microscopy, scanning electron microscopy (SEM), and Raman spectroscopy. Figures~\ref{fig:fig2}(a) and \ref{fig:fig2}(b) show the optical and SEM images of the sample, respectively. The SEM image reveals the presence of grain boundaries and structural defects, which are commonly observed in polycrystalline graphene films reported in earlier studies~\cite{ref31,ref39}. Two representative regions, marked as 1 (blue) and 2 (red), were selected from the SEM micrograph. Region~1 corresponds to an area close to a grain boundary, whereas Region~2 represents the interior of a graphene grain.
	
	Raman spectroscopy measurements performed at these two positions confirm their distinct structural characteristics, as shown in Figs.~\ref{fig:fig2}(c) and \ref{fig:fig2}(d). For the grain-boundary region (Region~1), the extracted intensity ratio is $I_{2D}/I_G \approx 1.4$, where $I_{2D}$ and $I_G$ represent the intensities of the 2D and G peaks, respectively. The peak positions and full width at half maximum (FWHM) values are $2684.30~\text{cm}^{-1}$ (45~$\text{cm}^{-1}$) for the 2D peak, $1584.067~\text{cm}^{-1}$ (30~$\text{cm}^{-1}$) for the G peak, and $1346.6~\text{cm}^{-1}$ (38~$\text{cm}^{-1}$) for the D peak. In contrast, the interior of the grain (Region~2) exhibits $I_{2D}/I_G \approx 2.1$, with corresponding peak positions and linewidths of $2683.08~\text{cm}^{-1}$ (37~$\text{cm}^{-1}$) for the 2D peak, $1583.08~\text{cm}^{-1}$ (27~$\text{cm}^{-1}$) for the G peak, and $1345.63~\text{cm}^{-1}$ (34~$\text{cm}^{-1}$) for the D peak. The reduced $I_{2D}/I_G$ ratio and increased linewidths at the grain boundary indicate enhanced defect-induced scattering and greater structural inhomogeneity. The broader 2D and G peaks further suggest increased local strain variations and possible doping fluctuations in this region. In contrast, the narrower linewidths and higher $I_{2D}/I_G$ ratio within the grain interior are consistent with improved crystalline uniformity and a lower defect density. These quantitative Raman parameters confirm the spatial variation in structural quality across the CVD-grown graphene film. Similar Raman signatures at grain boundaries in polycrystalline CVD graphene have also been reported in earlier studies~\cite{ref31,ref39}.

\begin{figure}[H]
	\includegraphics[width=0.48\textwidth]{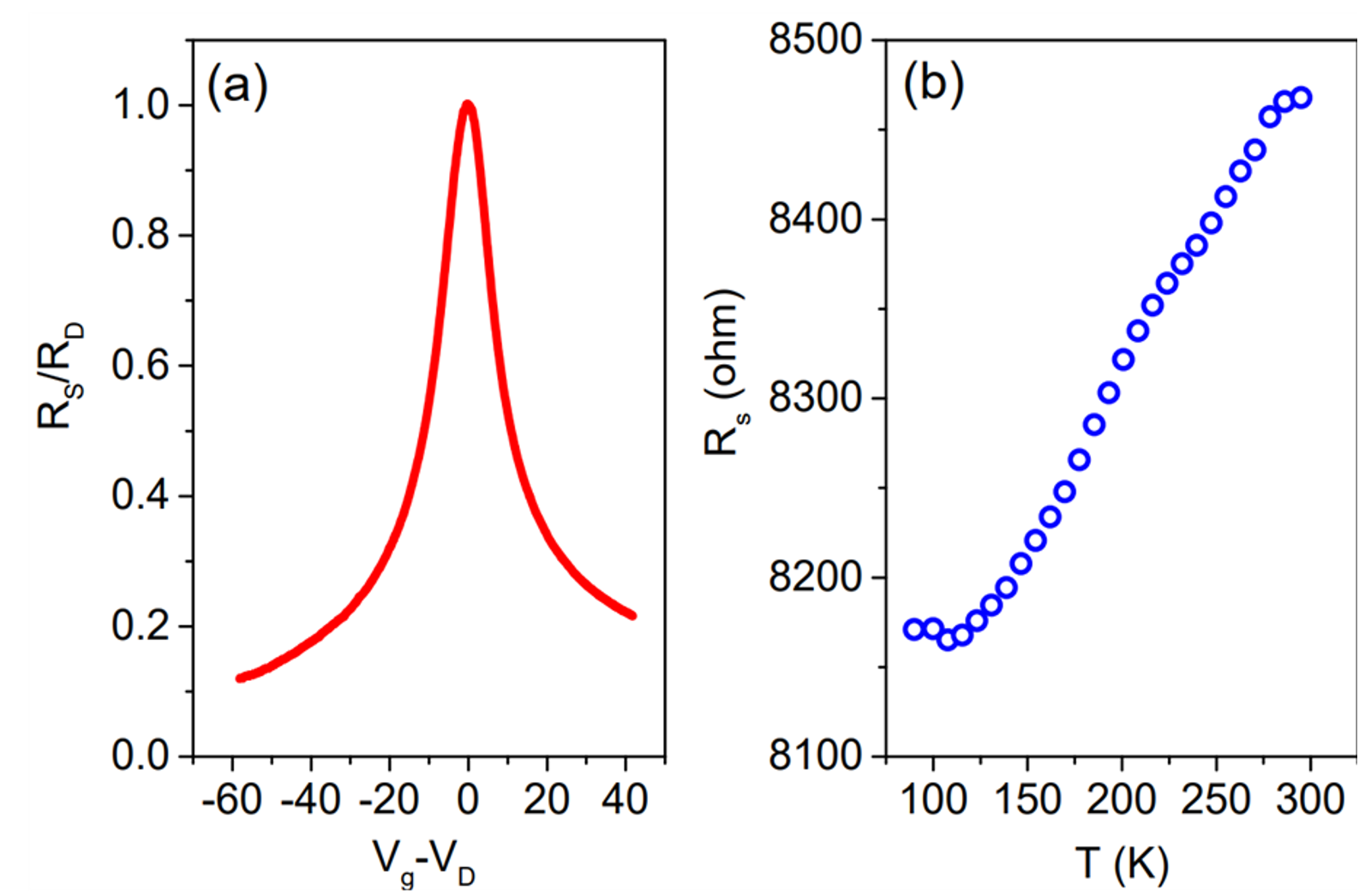}
	\caption{(a) Normalized sheet resistance of CVD-grown graphene as a function of gate voltage $(V_g - V_D)$ at $T = 80$~K. The Dirac point occurs at $V_D = 58$~V. (b) Temperature dependence of the sheet resistance measured at a fixed gate voltage $(V_g - V_D) = 10$~V.}
	\label{fig:fig3}
\end{figure}

From the combined optical, SEM, and Raman studies, we confirm that the CVD-grown graphene contains structural disorder such as grain boundaries and surface voids. To understand how these structural features influence electronic transport, we measured the resistance and resistance fluctuations of the sample over a wide temperature range from 295~K down to 80~K. Figure~\ref{fig:fig3}(a) presents the normalized sheet resistance $R_S$ as a function of reduced gate voltage $(V_g - V_D)$ measured at $T = 80$~K, where $V_D$ denotes the Dirac point of graphene. For our device, the Dirac point occurs at $V_D \approx 58$~V. Such a large positive shift indicates significant hole doping in the graphene channel. This shift is commonly attributed to adsorption of atmospheric species such as H$_2$O and O$_2$, which act as hole dopants and move the charge neutrality point toward positive gate voltages~\cite{ref17,ref18,ref40}. Figure~\ref{fig:fig3}(b) shows the temperature dependence of resistance at a fixed gate voltage $(V_g - V_D) = 10$~V. When the chemical potential is tuned away from the charge neutrality point, the resistance exhibits a positive temperature coefficient ($dR/dT > 0$) across the entire measurement range, similar to behavior reported in exfoliated graphene samples~\cite{ref41}. The extracted residual resistivity ratio (RRR), defined as $
	RRR = \frac{R_S(295\,\mathrm{K})}{R_S(80\,\mathrm{K})}
$, is found to be $RRR = 1.036$. Such a low value indicates significant disorder and defect-induced scattering within the sample, consistent with the structural characterization discussed earlier.

\begin{figure}[t]
	\includegraphics[width=0.48\textwidth]{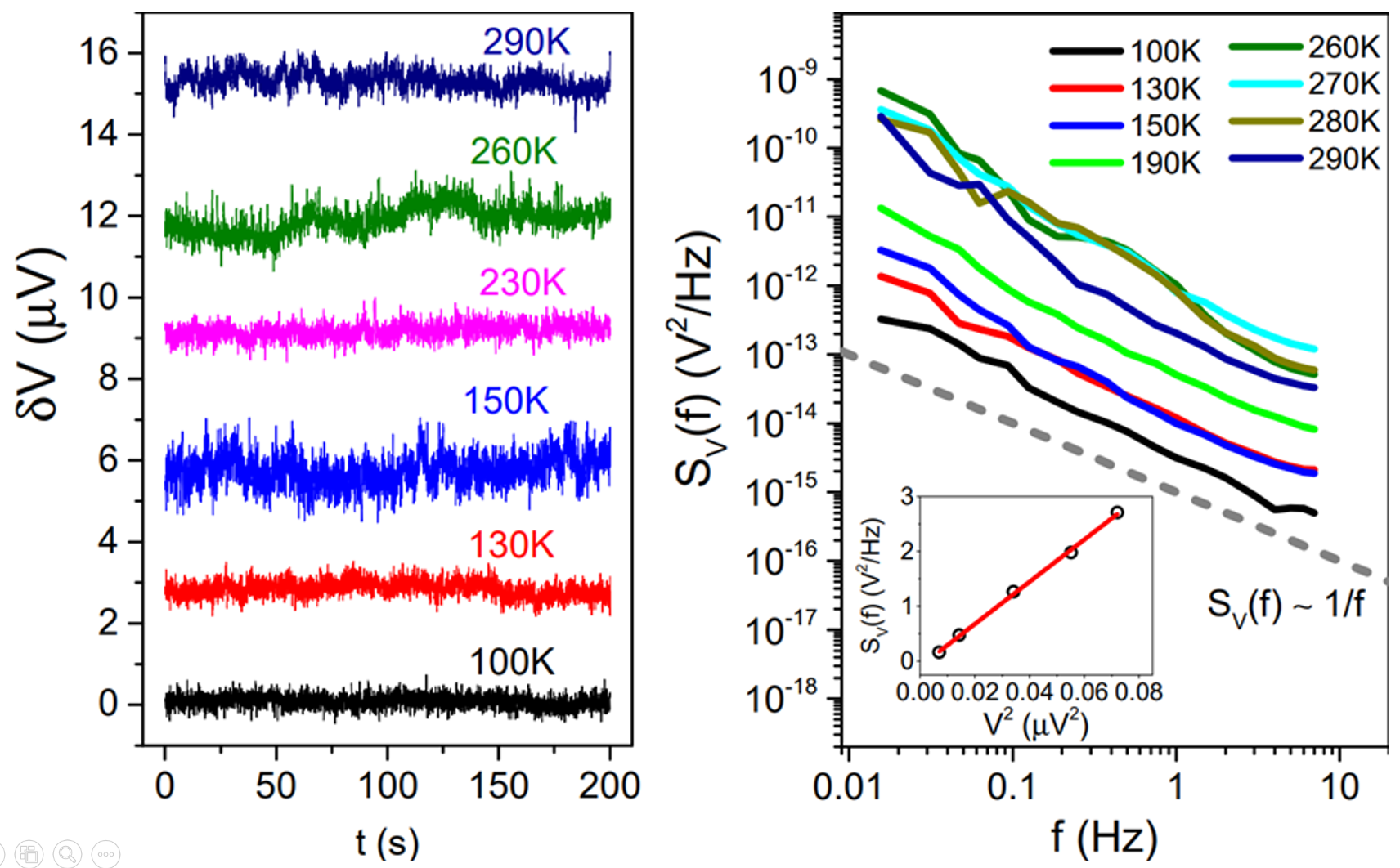}
	\caption{(a) Time series of voltage fluctuations measured at different temperatures at $(V_g - V_D) = 10$~V. The traces are vertically offset for clarity (offset value $3\,\mu$V). (b) Power spectral density of voltage fluctuations as a function of frequency. The dashed line represents a typical $1/f$ dependence. The inset shows the variance at $T=100$~K as a function of $V^2$, demonstrating quadratic scaling.}
	\label{fig:fig4}
\end{figure}

To investigate the influence of microstructural disorder on electronic transport, we measured low-frequency resistance fluctuations (noise) using a digital signal processing (DSP)-based AC four-probe technique~\cite{ref42,ref43}. Details of the measurement configuration and data analysis procedures are described in our earlier works~\cite{ref44,ref45,ref46,ref47,ref48,ref49}. This technique allows simultaneous detection of the intrinsic noise from the sample and the system background. A low-noise preamplifier (SR-552) was used to interface the sample with a lock-in amplifier. The excitation frequency of the lock-in amplifier was chosen within the noise figure minimum of the preamplifier to minimize amplifier-induced noise. The output signal from the lock-in amplifier was digitized using a high-speed 16-bit analog-to-digital converter and recorded for further analysis. Representative time series of voltage fluctuations at different temperatures are shown in Fig.~\ref{fig:fig4}(a). Each measurement run typically contained approximately $1.5 \times 10^{6}$ data points. The data were digitally filtered and decimated to remove 50~Hz line noise components. The filtered time series was used to compute the power spectral density (PSD) of voltage fluctuations $S_V(f)$ over a defined frequency window. The lower frequency cutoff (15~mHz) was limited by the stability of the temperature control system, which was better than $\pm1$~mK. The upper cutoff (7~Hz) was determined by the response limit of the lock-in amplifier output filter configured with a 10~ms time constant and 24~dB/octave roll-off. To ensure measurement accuracy, the setup was calibrated using the Johnson--Nyquist thermal noise of precision resistors, allowing reliable spectral measurements down to $S_V \sim 10^{-20}~\mathrm{V^2Hz^{-1}}$. The background noise floor exhibited no current dependence and matched the expected value of $4k_BTR_S$, confirming its Johnson noise origin.

The measured voltage spectral density was converted to resistance fluctuation spectral density using$	S_R(f) = \frac{S_V(f)}{I^2},$ where $I$ is the rms bias current applied to the sample. At all temperatures the spectra follow$	S_R(f) \propto \frac{1}{f^{\alpha}}$ with $\alpha$ close to unity (Fig.~\ref{fig:fig4}(b)). Furthermore, $S_V(f)$ exhibits a quadratic dependence on applied voltage, confirming that the observed $1/f$ noise originates from intrinsic resistance fluctuations of the sample. The values of $\alpha$, obtained from the slopes of the power spectral density (PSD) according to $\alpha = -\left(\partial \ln S_R(f) / \partial \ln f\right)$, are plotted in Fig.~6(b). We find that $\alpha$ remains nearly temperature independent down to 80~K, suggesting that the underlying noise mechanism does not change across the measured temperature range.

To compare the noise magnitude with other disordered electronic systems, we evaluate the Hooge parameter $\gamma_H$, defined as

\begin{equation}
	\gamma_H = N \frac{f S_V(f)}{V^2},
\end{equation}

where $N$ is the total number of charge carriers in the system. The extracted value at $T=100$~K and $(V_g - V_D)=10$~V is $\gamma_H \approx 5 \times 10^{2}$. This value is considerably larger than those reported for conventional crystalline metals and bulk semiconductors~\cite{ref50}, as well as exfoliated graphene~\cite{ref25}. However, it is comparable to values observed in oxide heterostructures such as the LaAlO$_3$/SrTiO$_3$ interface~\cite{ref45}, and significantly smaller than those reported in strongly disordered systems such as NbN thin films ($\gamma_H \sim 10^{5}$)~\cite{ref51} and rare-earth nickelate heterostructures ($\gamma_H \sim 10^{4}$)~\cite{ref49}.

\begin{figure}[H]
	\centering\includegraphics[width=0.45\textwidth]{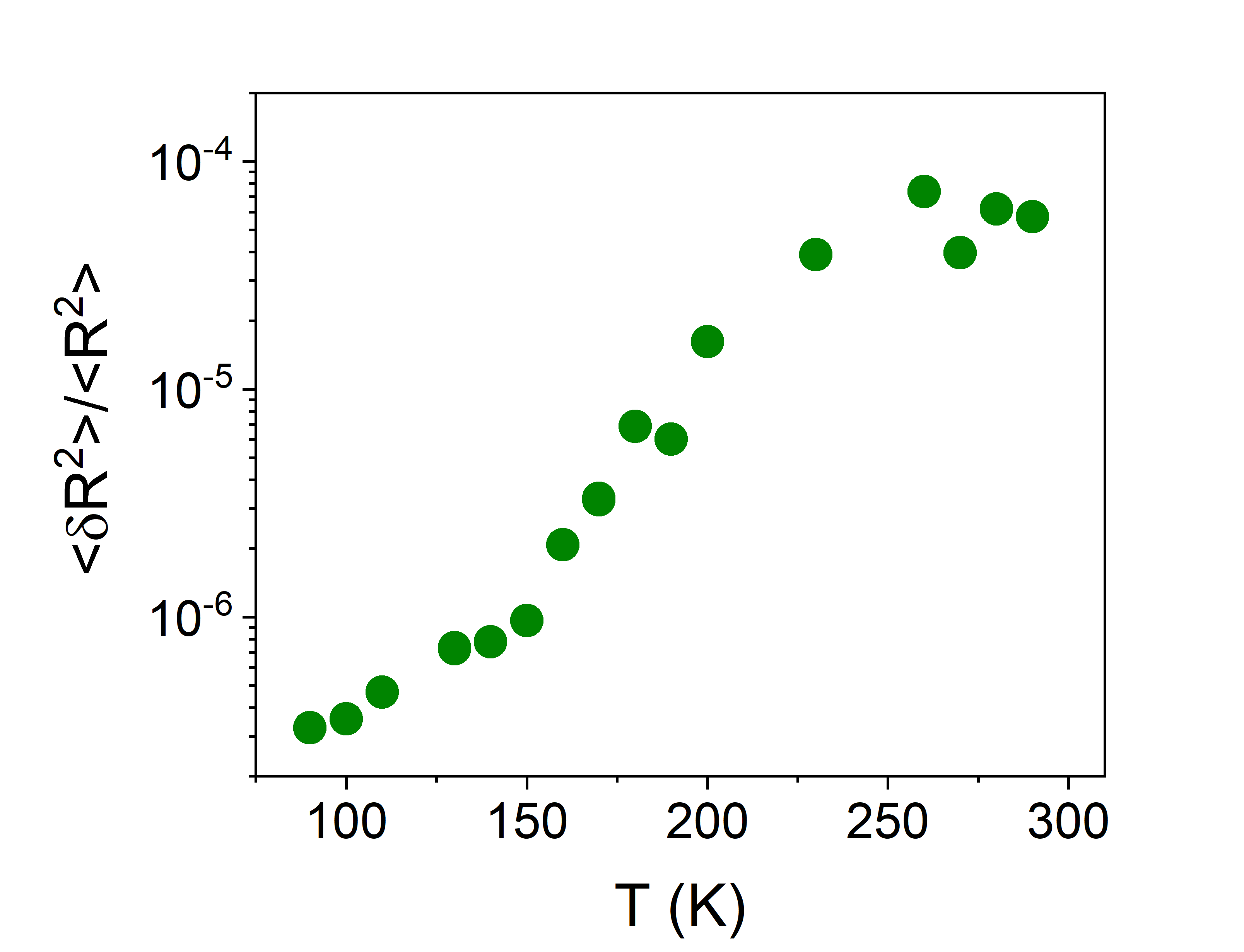}
	\caption{Relative variance of resistance fluctuations $\langle \delta R^2 \rangle / \langle R^2 \rangle$ as a function of temperature at $(V_g - V_D) = 10$~V.}
	\label{fig:fig5}
\end{figure}

A quantitative understanding of how resistance fluctuations evolve with temperature can be obtained by calculating the relative variance of the resistance noise. The relative variance of resistance fluctuations is defined as

\begin{equation}
	\frac{\langle \delta R^2 \rangle}{\langle R^2 \rangle}
	=
	\frac{1}{V^2}
	\int_{15\,\text{mHz}}^{7\,\text{Hz}} S_V(f)\,df.
\end{equation}

where $\langle \delta R^2 \rangle$ represents the variance of the sheet resistance and $V$ is the voltage across the device. Figure~5 shows the temperature dependence of the relative variance of resistance fluctuations, $\langle \delta R^2 \rangle / \langle R^2 \rangle$, measured at a fixed reduced gate voltage of $V_g - V_D = 10$~V. In continuous single-component metallic films, low-frequency resistance noise is generally attributed to local fluctuations in the resistance arising from variations in scattering rates caused by mobile defects, impurities, or other slow structural rearrangements that influence charge carrier motion. In such systems, the magnitude of the noise typically scales with the overall resistivity of the film~\cite{ref52}. Our observations are consistent with this behavior. As the temperature increases toward room temperature, the relative variance of resistance fluctuations increases substantially, with the room-temperature noise level nearly two orders of magnitude larger than that measured at the lowest temperatures. This pronounced temperature dependence indicates that thermally activated defect dynamics play a dominant role in governing resistance fluctuations in the CVD-grown graphene sample. The strong temperature dependence of the resistance noise can be understood within the framework of the Dutta--Horn model, which describes $1/f$ noise as arising from a superposition of a large number of thermally activated fluctuators, each characterized by a relaxation time $\tau_0$~\cite{ref53}. Within this model, the distribution $D(E_0)$ of activation energies $E_0$ associated with these fluctuators can be extracted from the measured power spectral density of resistance fluctuations according to~\cite{ref53,ref54,ref55}:
\begin{equation}
	D(E_0) \propto \frac{\omega S_V(\omega,T)}{k_B T},
\end{equation}
where $\omega = 2\pi f$ is the angular frequency of the measurement and $S_V(\omega,T)$ is the power spectral density (PSD) of the voltage fluctuations. The activation energy is given by $E_0 = k_B T \ln(2\pi f \tau_0)$, where $\tau_0$ represents the attempt time of the thermally activated process. Typical values of $\tau_0$ correspond to the intrinsic attempt time of the fluctuator and are generally of the order of $10^{-14}$~s, which is comparable to the inverse of a characteristic phonon frequency~\cite{ref53}. 
\begin{figure}[t]
	\centering
	
	\subfloat{
		\centering\includegraphics[width=0.7\columnwidth]{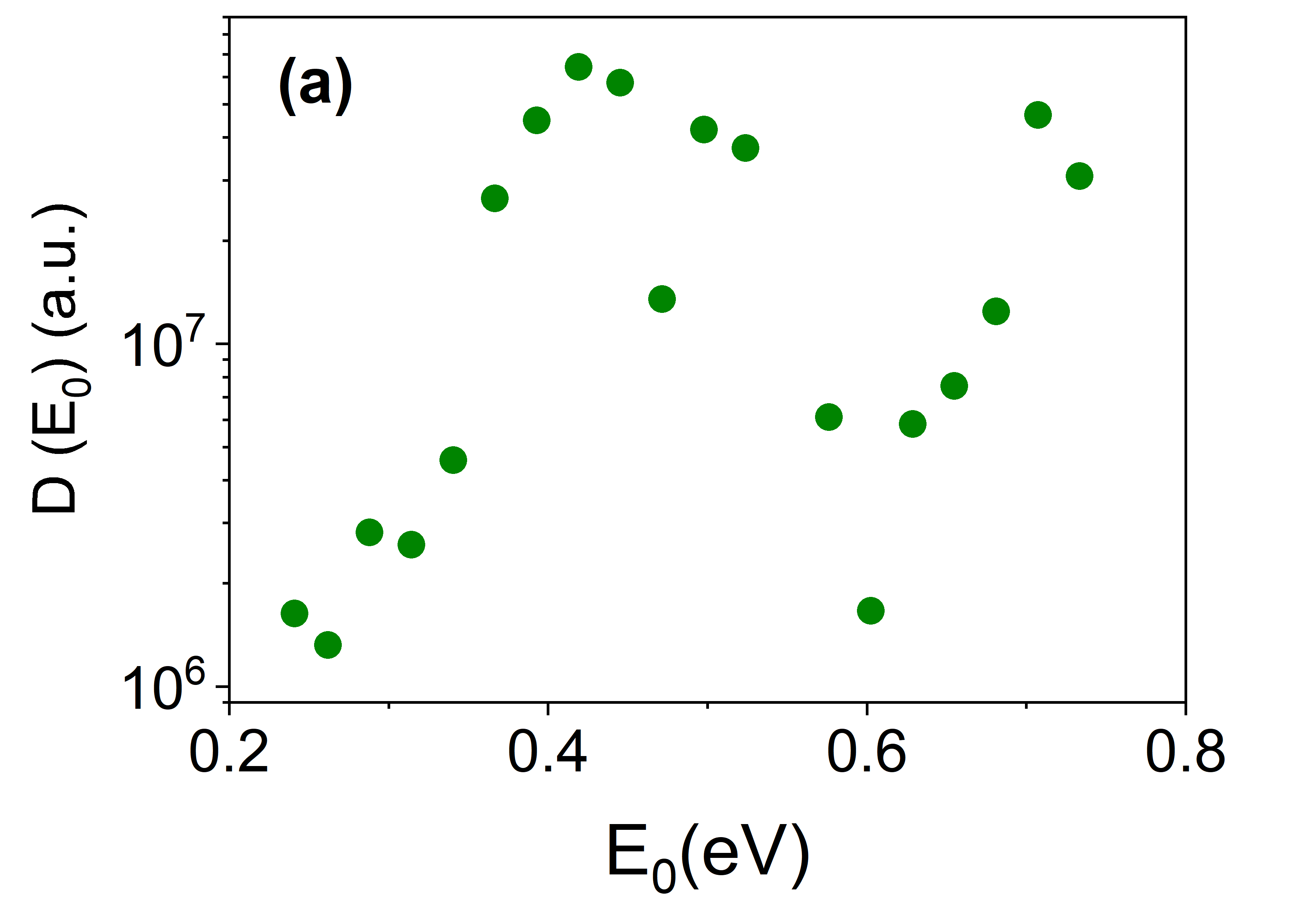}
	}
	
	\vspace{0.5cm}
	
	\subfloat{
		\centering\includegraphics[width=0.6\columnwidth]{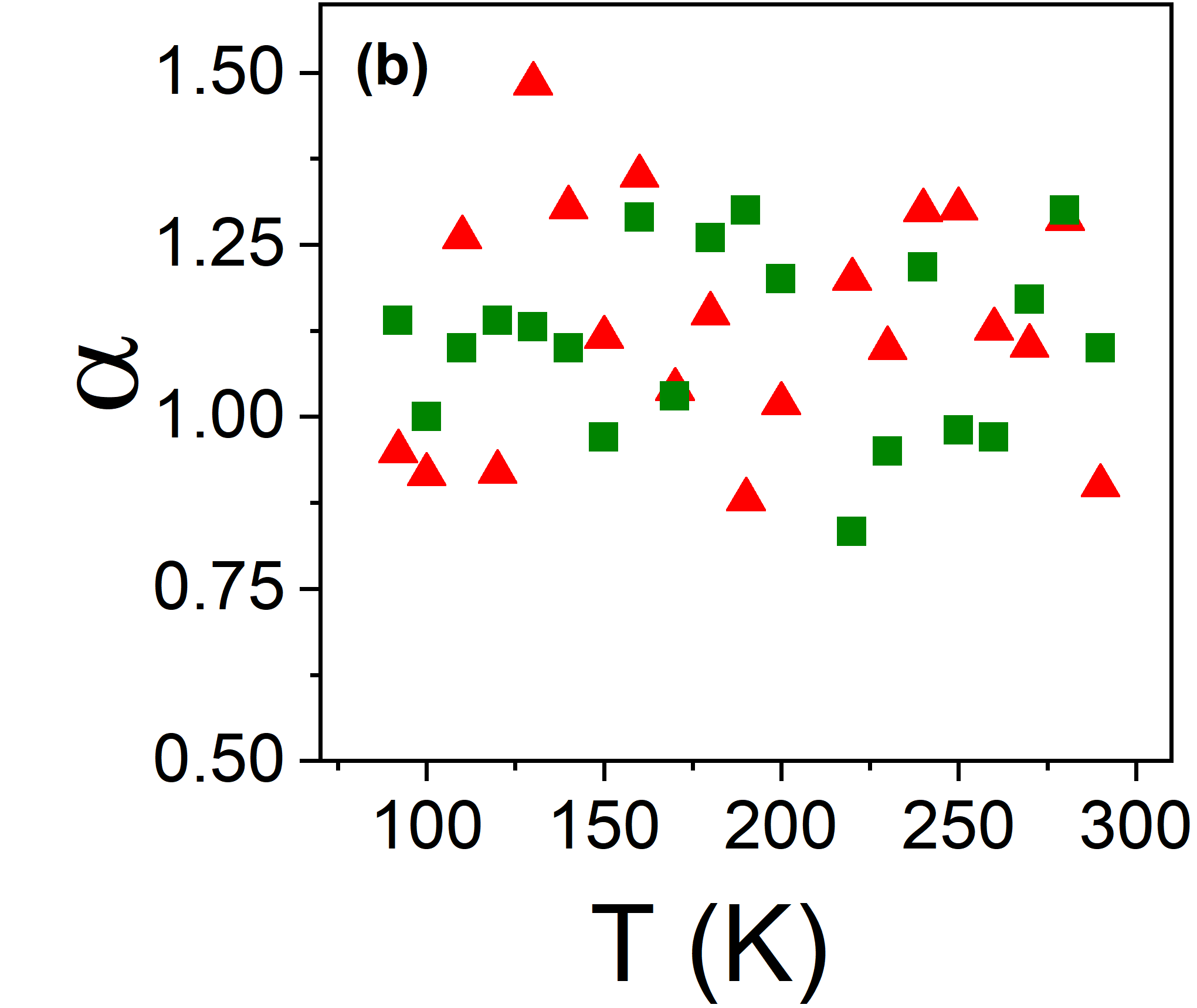}
	}
	
	\caption{(a) Energy dependence of the noise amplitude parameter $D(E_0)$ as a function of Fermi energy. (b) Temperature dependence of the noise exponent $\alpha$. Red triangles correspond to values obtained using the Dutta–Horn model, while olive squares represent values extracted from the slope of the PSD.}
	
	\label{fig:fig6}
\end{figure}

Using the above relation, the distribution $D(E_0)$ is calculated from the experimentally measured noise spectra and is plotted in Fig.~6(a). The resulting curve represents the variation of $D(E_0)$ as a function of activation energy $E_0$, providing information about the density and strength of the fluctuators responsible for the observed $1/f$ noise.If the resistance fluctuations indeed arise from thermally activated dynamics of local defects, the frequency exponent $\alpha$ can also be determined independently from the temperature dependence of the noise within the framework of the Dutta--Horn (DH) model~\cite{ref54}. According to this model, the temperature dependence of $\alpha$ is given by

\begin{equation}
	\alpha(\omega,T)
	=
	1
	-
	\frac{1}{\ln(\omega\tau_0)}
	\left(
	\frac{\partial \ln S_R(\omega,T)}{\partial \ln T} - 1
	\right).
\end{equation}

where $\omega = 2\pi f$ and $S_R(\omega,T)$ is the power spectral density (PSD) of the resistance fluctuations. Figure~6(b) shows the values of $\alpha$ extracted from this expression as a function of temperature for $V_g - V_D = 10$~V (red filled triangles). These values are in close agreement with those obtained independently from the slope of the measured PSD of resistance fluctuations, where $\alpha = -\left(\partial \ln S_R(f) / \partial \ln f \right)$ (olive filled squares). We note that grain boundaries are expected to be the dominant sources of noise in the present system. The localized electronic states associated with grain boundaries can trap and release charge carriers, leading to fluctuations in carrier density as well as variations in the effective potential barriers for charge transport~\cite{ref56}. Furthermore, grain boundaries are chemically more reactive due to the presence of dangling bonds and local strain, making them favorable sites for the adsorption of environmental molecules, which can further enhance slow charge fluctuations. In addition to grain boundaries, charge traps in the SiO$_2$ substrate can also contribute to the observed noise through electrostatic coupling with the graphene channel. Other structural defects, such as vacancies and ripples, may also contribute to resistance fluctuations; however, their influence is expected to be comparatively weaker owing to their lower density~\cite{ref21,ref31}. The excellent agreement between the experimentally extracted values of $\alpha$ and the predictions of the Dutta--Horn model strongly supports the conclusion that the dominant source of resistance fluctuations in CVD-grown graphene originates from thermally activated switching of local defect-induced scatterers.

\section{Conclusion}
In summary, we have systematically investigated the temperature dependence of resistance and low-frequency 1/f noise in CVD-grown graphene device. The measurements demonstrate that the observed resistance fluctuations originate from thermally activated fluctuations of local defects. These findings highlight the crucial role of defect dynamics in governing electronic noise, particularly at elevated temperatures. Future studies may explore growth-stage control to suppress grain boundary formation, improved transfer-process engineering to minimize residue-induced and mechanical defects, and interface engineering through substrate selection, surface treatment, and encapsulation to reduce electrically active trap densities. Such strategies could provide a practical route toward defect engineering for the scalable fabrication of low-noise, high-performance CVD-grown graphene devices. Our results provide important insights for optimizing material quality and improving the performance and reliability of CVD grown graphene in temperature-sensitive electronic applications.

\begin{acknowledgments}
G.N.D. acknowledges support from the Anusandhan National Research Foundation (ANRF) under IRG grant (Grant No. ANRF/IRG/2024/000714/PS) and the ANRF PMECRG (Grant No. ANRF/ECRG/2024/001810/PMS), Government of India. G.N.D. also acknowledges the start-up research grant from NIT Rourkela.
\end{acknowledgments}
	
	\section*{Data Availability}
	
	The data that support the findings of this study are available from the corresponding author upon reasonable request.
	
\bibliographystyle{apsrev4-2}
\bibliography{Graphene_Noise}

\begin{thebibliography}{56}
	\bibitem{ref1}
	A.K. Geim and K.S. Novoselov,
	“The rise of graphene,”
	Nature Materials 6, no. 3 (2007): 183–191.
	doi:10.1038/nmat1849.
	
	\bibitem{ref2}
	Y. Zheng and T. Ando,
	“Hall conductivity of a two-dimensional graphite system,”
	Physical Review B 65, no. 24 (2002): 245420.
	doi:10.1103/PhysRevB.65.245420.
	
	\bibitem{ref3}
	V.P. Gusynin and S.G. Sharapov,
	“Unconventional Integer Quantum Hall Effect in Graphene,”
	Physical Review Letters 95, no. 14 (2005): 146801.
	doi:10.1103/PhysRevLett.95.146801.
	
	\bibitem{ref4}
	A.H. Castro Neto, F. Guinea, N.M.R. Peres, K.S. Novoselov and A.K. Geim,
	“The electronic properties of graphene,”
	Reviews of Modern Physics 81, no. 1 (2009): 109–162.
	doi:10.1103/RevModPhys.81.109.
	
	\bibitem{ref5}
	Y. Zhang, Y.-W. Tan, H.L. Stormer and P. Kim,
	“Experimental observation of the quantum Hall effect and Berry's phase in graphene,”
	Nature 438, no. 7065 (2005): 201–204.
	doi:10.1038/nature04235.
	
	\bibitem{ref6}
	K.I. Bolotin, K.J. Sikes, J. Hone, H.L. Stormer and P. Kim,
	“Temperature-Dependent Transport in Suspended Graphene,”
	Physical Review Letters 101, no. 9 (2008): 096802.
	doi:10.1103/PhysRevLett.101.096802.
	
	\bibitem{ref7}
	K. von Klitzing, G. Dorda and M. Pepper,
	“New Method for High-Accuracy Determination of the Fine-Structure Constant Based on Quantized Hall Resistance,”
	Physical Review Letters 45, no. 6 (1980): 494–497.
	doi:10.1103/PhysRevLett.45.494.
	
	\bibitem{ref8}
	A.F. Young et al.,
	“Spin and valley quantum Hall ferromagnetism in graphene,”
	Nature Physics 8, no. 7 (2012): 550–556.
	doi:10.1038/nphys2307.
	
	\bibitem{ref9}
	X. Li et al.,
	“Graphene films with large domain size by a two-step chemical vapor deposition process,”
	Nano Letters 10, no. 11 (2010): 4328–4334.
	
	\bibitem{ref10}
	H. Zhou et al.,
	“Chemical vapour deposition growth of large single crystals of monolayer and bilayer graphene,”
	Nature Communications 4 (2013): 2096.
	doi:10.1038/ncomms3096.
	
	\bibitem{ref11}
	H. Zhang et al.,
	“Grain Boundary Effect on Electrical Transport Properties of Graphene,”
	The Journal of Physical Chemistry C 118, no. 5 (2014): 2338–2343.
	doi:10.1021/jp411464w.
	
	\bibitem{ref12}
	Q. Yu et al.,
	“Control and characterization of individual grains and grain boundaries in graphene grown by chemical vapour deposition,”
	Nature Materials 10, no. 6 (2011): 443–449.
	doi:10.1038/nmat3010.
	
	\bibitem{ref13}
	Z. Fei et al.,
	“Electronic and plasmonic phenomena at graphene grain boundaries,”
	Nature Nanotechnology 8, no. 11 (2013): 821–825.
	doi:10.1038/nnano.2013.197.
	
	\bibitem{ref14}
	H. Cao et al.,
	“Electronic transport in chemical vapor deposited graphene synthesized on Cu: Quantum Hall effect and weak localization,”
	Applied Physics Letters 96 (2010): 122106.
	
	\bibitem{ref15}
	A. Mesaros et al.,
	“Electronic states of graphene grain boundaries,”
	Physical Review B 82 (2010): 205119.
	doi:10.1103/PhysRevB.82.205119.
	
	\bibitem{ref16}
	O.V. Yazyev and S.G. Louie,
	“Electronic transport in polycrystalline graphene,”
	Nature Materials 9 (2010): 806–809.
	doi:10.1038/nmat2830.
	
	\bibitem{ref17}
	G.N. Daptary et al.,
	“Enhancement of superconductivity upon reduction of carrier density in proximitized graphene,”
	Physical Review B 105 (2022): L100507.
	doi:10.1103/PhysRevB.105.L100507.
	
	\bibitem{ref18}
	G.N. Daptary et al.,
	“Observation of a second Dirac point in a graphene/superconductor bilayer,”
	Physical Review Materials 8 (2024): 084802.
	doi:10.1103/PhysRevMaterials.8.084802.
	
	\bibitem{ref19}
	J. Glemža et al.,
	“Low-frequency noise characteristics of graphene/h-BN/Si junctions,”
	Crystals 15 (2025): 747.
	
	\bibitem{ref20}
	A. Rehman et al.,
	“Nature of the 1/f noise in graphene—direct evidence for the mobility fluctuation mechanism,”
	Nanoscale 14 (2022): 7242–7249.
	
	\bibitem{ref21}
	V. Kochat et al.,
	“Magnitude and origin of electrical noise at individual grain boundaries in graphene,”
	Nano Letters 16 (2016): 562–567.
	
	\bibitem{ref22}
	M. Tian et al.,
	“Tunable 1/f noise in CVD Bernal-stacked bilayer graphene transistors,”
	ACS Applied Materials \& Interfaces 12 (2020): 17686–17690.
	
	\bibitem{ref23}
	A. Mehra, R.J. Mathew and C. Kumar,
	“Origin of electrical noise near charge neutrality in dual gated graphene device,”
	Applied Physics Letters 123 (2023): 123103.
	
	\bibitem{ref24}
	C. Kumar and A. Das,
	“Effect of boron nitride defects and charge inhomogeneity on 1/f noise in encapsulated graphene,”
	Applied Physics Letters 119 (2021).
	doi:10.1063/5.0071152.
	
	\bibitem{ref25}
	A.A. Balandin,
	“Low-frequency 1/f noise in graphene devices,”
	Nature Nanotechnology 8 (2013): 549–555.
	
	\bibitem{ref26}
	A.A. Balandin, E. Paladino and P.J. Hakonen,
	“Electronic noise—From advanced materials to quantum technologies,”
	Applied Physics Letters 124 (2024): 050401.
	
	\bibitem{ref27}
	K.A. Kazakov,
	“The fundamental 1/f noise in monolayer graphene,”
	International Journal of Modern Physics B 38 (2024): 2450138.
	
	\bibitem{ref28}
	M. Kamada et al.,
	“Suppression of 1/f noise in graphene due to anisotropic mobility fluctuations induced by impurity motion,”
	Communications Physics 6 (2023): 207.
	
	\bibitem{ref29}
	M. Kumar et al.,
	“Ultra low 1/f noise in suspended bilayer graphene,”
	Applied Physics Letters 106 (2015).
	
	\bibitem{ref30}
	V.K. Sangwan et al.,
	“Low-frequency electronic noise in single-layer MoS2 transistors,”
	Nano Letters 13 (2013): 4351–4355.
	
	\bibitem{ref31}
	V. Kochat et al.,
	“Origin of 1/f noise in graphene produced for large-scale applications in electronics,”
	IET Circuits, Devices \& Systems 9 (2015): 52–58.
	
	\bibitem{ref32}
	A. Schmitt et al.,
	“High-field 1/f noise in hBN-encapsulated graphene transistors,”
	Physical Review B 107 (2023): L161104.
	
	\bibitem{ref33}
	B. Zhang et al.,
	“Analysis of low-frequency 1/f noise characteristics for MoTe2 ambipolar field-effect transistors,”
	Nanomaterials 12 (2022): 1325.
	
	\bibitem{ref34}
	S. Ghosh et al.,
	“Low‐Frequency Current Fluctuations in Quasi‐1D (TaSe4)2I Weyl Semimetal Nanoribbons,”
	Advanced Electronic Materials 9 (2023): 2200860.
	
	\bibitem{ref35}
	Y. Kim et al.,
	“Low-frequency noise behaviors of quasi-two-dimensional electron systems based on complex oxide heterostructures,”
	Current Applied Physics 59 (2024): 129–135.
	
	\bibitem{ref36}
	G.N. Daptary et al.,
	“Effect of multiband transport on charge carrier density fluctuations at the LaAlO3/SrTiO3 interface,”
	Physical Review B 98 (2018): 035433.
	
	\bibitem{ref37}
	G.N. Daptary et al.,
	“Correlated non-Gaussian phase fluctuations in LaAlO3/SrTiO3 heterointerfaces,”
	Physical Review B 94 (2016): 085104.
	
	\bibitem{ref38}
	G.N. Daptary et al.,
	“Effect of spin-orbit interaction on the vortex dynamics in LaAlO3/SrTiO3 interfaces near the superconducting transition,”
	Physical Review B 100 (2019): 125117.
	
	\bibitem{ref39}
	C. Bautista-Flores et al.,
	“Raman spectroscopy of CVD graphene during transfer process from copper to SiO2/Si substrates,”
	Materials Research Express 6 (2019): 015601.
	
	\bibitem{ref40}
	Y.J. Shin et al.,
	“Surface-Energy Engineering of Graphene,”
	Langmuir 26 (2010): 3798–3802.
	
	\bibitem{ref41}
	S. Sarkar et al.,
	“Role of different scattering mechanisms on the temperature dependence of transport in graphene,”
	Scientific Reports 5 (2015): 16772.
	
	\bibitem{ref42}
	J.H. Scofield,
	“AC method for measuring low-frequency resistance fluctuation spectra,”
	Review of Scientific Instruments 58 (1987): 985–993.
	
	\bibitem{ref43}
	A. Ghosh et al.,
	“A set-up for measurement of low frequency conductance fluctuation (noise) using digital signal processing techniques,”
	cond-mat/0402130 (2004).
	
	\bibitem{ref44}
	G.N. Daptary et al.,
	Physical Review B 100 (2019): 125117.
	
	\bibitem{ref45}
	G.N. Daptary et al.,
	Physical Review B 98 (2018): 035433.
	
	\bibitem{ref46}
	G.N. Daptary et al.,
	Physical Review B 94 (2016): 085104.
	
	\bibitem{ref47}
	G.N. Daptary et al.,
	“Probing a spin-glass state in SrRuO3 thin films through higher-order statistics of resistance fluctuations,”
	Physical Review B 90 (2014): 115153.
	
	\bibitem{ref48}
	G.N. Daptary et al.,
	“Effect of microstructure on the electronic transport properties of epitaxial CaRuO3 thin films,”
	Physica B 511 (2017): 74–79.
	
	\bibitem{ref49}
	G.N. Daptary et al.,
	“Conductivity noise across temperature-driven transitions of rare-earth nickelate heterostructures,”
	Physical Review B 100 (2019): 125105.
	
	\bibitem{ref50}
	F.N. Hooge,
	“1/f noise sources,”
	IEEE Transactions on Electron Devices 41 (1994): 1926–1935.
	
	\bibitem{ref51}
	R. Koushik et al.,
	“Correlated Conductance Fluctuations Close to the Berezinskii-Kosterlitz-Thouless Transition in Ultrathin NbN Films,”
	Physical Review Letters 111 (2013): 197001.
	
	\bibitem{ref52}
	J.H. Scofield, J.V. Mantese and W.W. Webb,
	“1/f noise of metals: A case for extrinsic origin,”
	Physical Review B 32 (1985): 736–742.
	
	\bibitem{ref53}
	P. Dutta, P. Dimon and P.M. Horn,
	“Energy Scales for Noise Processes in Metals,”
	Physical Review Letters 43 (1979): 646–649.
	
	\bibitem{ref54}
	P. Dutta and P.M. Horn,
	“Low-frequency fluctuations in solids: 1/f noise,”
	Reviews of Modern Physics 53 (1981): 497–516.
	
	\bibitem{ref55}
	D.M. Fleetwood and N. Giordano,
	“Direct link between 1/f noise and defects in metal films,”
	Physical Review B 31 (1985): 1157–1160.
	
	\bibitem{ref56}
	A.N. Pal, A.A. Bol and A. Ghosh,
	“Large low-frequency resistance noise in chemical vapor deposited graphene,”
	Applied Physics Letters 97 (2010).
\end{thebibliography}
	
\end{document}